\documentclass[journal]{IEEEtran}
\usepackage{amsmath,amssymb,amsfonts}
\usepackage{algorithmic}
\usepackage{graphicx}
\usepackage{textcomp}
\usepackage{xcolor}
\usepackage{cite}
\usepackage{hyperref}
\usepackage{tikz}
\usepackage{booktabs}

\begin{document}

\title{Uncertainty Quantification and Data Efficiency in AI: An Information-Theoretic Perspective}

\author{Osvaldo~Simeone and Yaniv Romano
\vspace{-0.5cm} 
\thanks{O. Simeone is with the Institute for Intelligent Networked Systems, Northeastern University London, One Portsoken Street, London, E1 8PH, UK (email: o.simeone@northeastern.edu).  Y. Romano is with the Departments of Electrical and Computer Engineering and of Computer Science at the Technion--Israel Institute of Technology, Haifa, Israel (email: yromano@technion.ac.il). The work of O. Simeone was supported by the European Research Council (ERC) under the European Union’s Horizon Europe Programme (grant agreement No. 101198347), by an Open Fellowship of the EPSRC (EP/W024101/1), and  by the EPSRC project  EP/X011852/1.
 The work of Y. Romano was supported by the by the European Union (ERC, SafetyBounds, 101163414). Views and opinions expressed are however those of the authors only and do not necessarily reflect those of the European Union or the European Research Council Executive Agency. Neither the European Union nor the granting authority can be held responsible for them.}
}

\maketitle

\begin{abstract}
In context-specific  applications such as robotics, telecommunications, and healthcare, artificial intelligence systems often face the challenge of limited training data.  This scarcity introduces epistemic uncertainty, i.e., reducible uncertainty stemming from incomplete knowledge of the underlying data distribution, which fundamentally limits predictive performance. This review paper examines formal methodologies that address data-limited regimes through two complementary approaches: quantifying epistemic uncertainty and mitigating data scarcity via synthetic data augmentation. We begin by reviewing generalized Bayesian learning frameworks that characterize epistemic uncertainty through generalized posteriors in the model parameter space, as well as ``post-Bayes'' learning frameworks. We continue by presenting  information-theoretic generalization bounds that formalize the relationship between training data quantity and predictive uncertainty, providing a theoretical justification for generalized Bayesian learning. Moving beyond methods with asymptotic statistical validity, we  survey uncertainty quantification methods that provide finite-sample statistical guarantees, including conformal prediction and conformal risk control. Finally, we examine recent advances in data efficiency by combining limited labeled data with abundant model predictions or synthetic data. Throughout, we take an information-theoretic perspective, highlighting the role of information measures in quantifying the impact of data scarcity. 
\end{abstract}


\section{Introduction}

The remarkable success of modern artificial intelligence (AI) systems based on large language models (LLMs) has been largely predicated on the availability of massive, Internet-scale training datasets. However, this data abundance does not translate to many critical application domains. In robotics, collecting diverse demonstrations requires expensive human expert time and controlled environments \cite{ankile2024juicer}. In telecommunications, optimal control policies are inherently deployment-specific, making pertinent data inherently scarce \cite{polese2024colosseum,ruah2025bridge}. In healthcare, patient-specific data is limited by privacy constraints, regulatory requirements, and the natural rarity of certain conditions \cite{kumari2023dataefficient}. For these high-stakes, data-scarce environments, the ultimate goal is personalization,  or   more broadly specialization, moving from predictions that are only useful on average to  predictions that are relevant for specific deployments, individuals,  or subpopulations.

\begin{figure*}[!t]
\centering
\begin{minipage}[b]{0.48\linewidth}
    \centering
    \includegraphics[width=\linewidth]{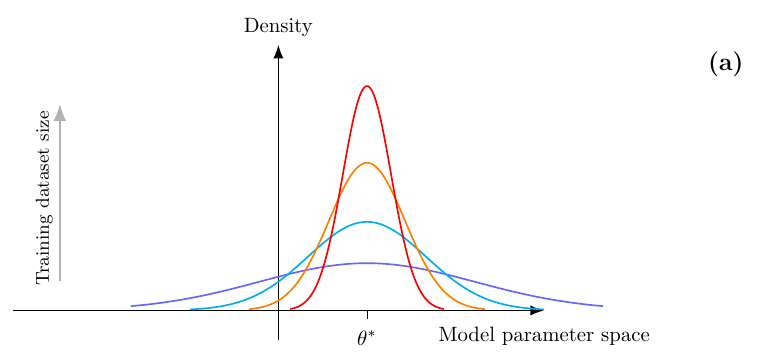}
\end{minipage}
\hfill
\begin{minipage}[b]{0.48\linewidth}
    \centering
    \includegraphics[width=\linewidth]{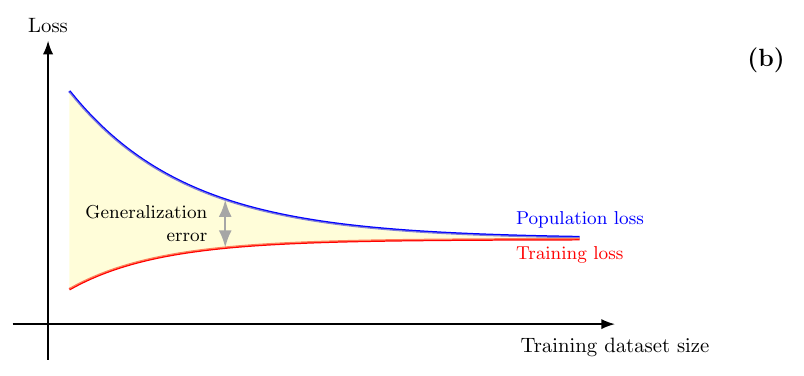}
\end{minipage}

\vspace{0.5cm}

\begin{minipage}[b]{0.48\linewidth}
    \centering
    \includegraphics[width=\linewidth]{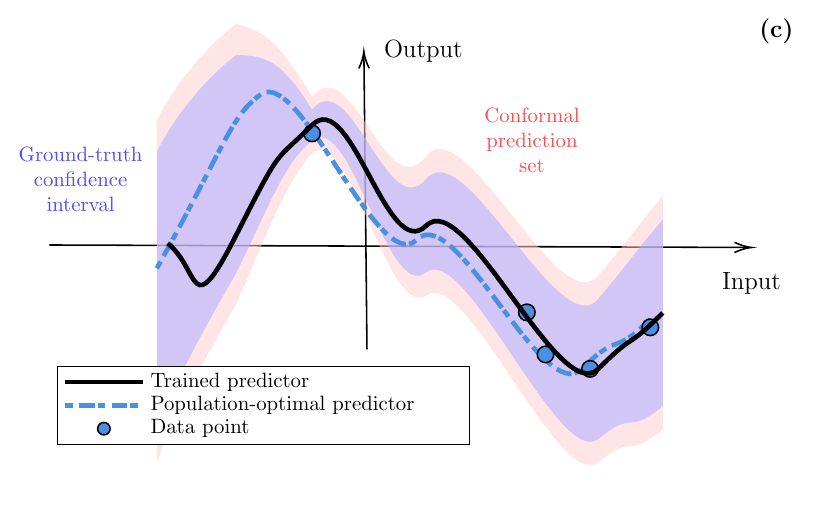}
\end{minipage}
\hfill
\begin{minipage}[b]{0.48\linewidth}
    \centering
    \includegraphics[width=\linewidth]{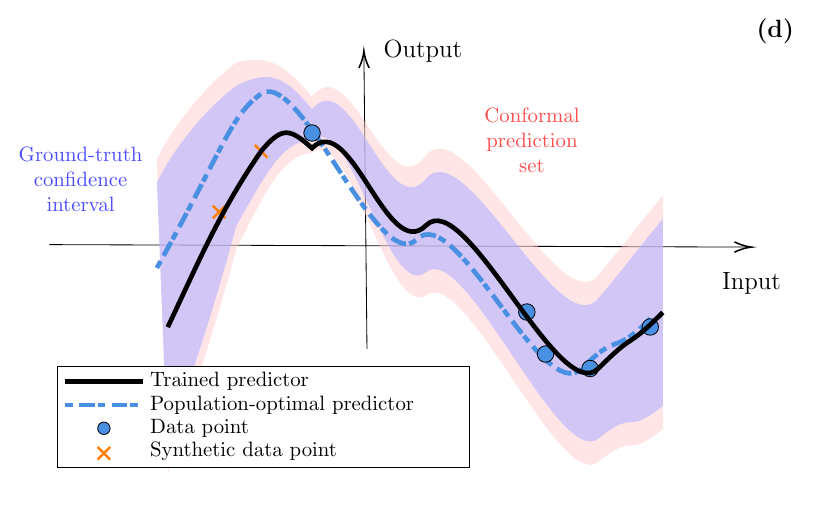}
\end{minipage}
\caption{Key concepts on uncertainty-aware data-efficient AI reviewed in this article: (a) Bayesian learning, and post-Bayes variants thereof, describe uncertainty in the model parameter space by relying on the specification of a prior distribution and a likelihood/loss function, or possibly a predictive distribution; (b) Generalization bounds provide insights into the gap between population loss (i.e., risk) and training loss as a function of the training dataset size; (c) Conformal prediction, and variants thereof, provide means to calibrate prediction sets (e.g., error bars) so that they contain the true output with a user-specified coverage probability; and (d) Synthetic data can be leveraged to both enhance the trained predictor and improve the quality of the prediction sets.}\label{fig:1}
\end{figure*}

Data scarcity fundamentally manifests as \textbf{epistemic uncertainty}. Epistemic uncertainty arises from incomplete knowledge of the true data-generating distribution \cite{hullermeier2021aleatoric,simeone2022machine}. Unlike aleatoric uncertainty, which captures irreducible randomness inherent to the data-generation mechanism, epistemic uncertainty can be reduced by acquiring more pertinent training data or by refining model specifications \cite{simeone2022machine}.

Uncertainty and information are two sides of the same coin. Information theory provides a rigorous mathematical framework for analyzing and quantifying uncertainty and information \cite{cover2006elements,simeone2024classical}. Information-theoretic measures, such as mutual information and relative entropy, have accordingly found extensive application as tools to study data-limited learning. For example, it was shown that the mutual information between model parameters and training data -- capturing the sensitivity of the trained model to data -- provides bounds on the generalization error of AI models \cite{hellstrom2023generalization,banchi2024statistical}. 


This review paper covers principles and recent advances on uncertainty-aware AI for data-limited settings from an information-theoretic perspective. The presentation is organized around two complementary strategies for data-efficient AI:

\begin{itemize}
\item \textbf{Quantifying epistemic and predictive uncertainty:} In the presence of limited training data, epistemic uncertainty fundamentally limits the precision of any AI model. Uncertainty quantification is thus an essential step to ensure the reliable deployment of AI models in domains such as robotics, telecommunications, and healthcare. We start by reviewing Bayesian learning as a formal framework to quantify epistemic uncertainty in the parameter space, as sketched in Figure \ref{fig:1}(a) (with $\theta$ representing the model parameters). We also cover recent ``post-Bayesian'' advances including generalized Bayesian learning \cite{knoblauch2022optimization} and martingale posteriors \cite{knoblauch2022optimization,simeone2022machine}. We then review information-theoretic generalization bounds \cite{hellstrom2023generalization}. As illustrated in Figure \ref{fig:1}(b), generalization bounds offer insights into the scaling of the average error caused by limitations in the training data. As it will be shown, information-theoretic generalization bounds provide a theoretical justification for the use of generalized Bayesian learning. However, these bounds do not provide any actionable quantification of uncertainty for individual inputs. As exemplified in Figure \ref{fig:1}(c), this type of information can be instead obtained via prediction sets calibrated through conformal prediction, which offer distribution-free, finite-sample  coverage guarantees \cite{vovk2005algorithmic,angelopoulos2024theoretical}.

\item \textbf{Leveraging synthetic data:} While uncertainty quantification is a required step whenever training data are in limited supply, in practice one may wish to find ways to augment the dataset with additional data. When real data cannot be collected, or is too expensive to obtain, the only remaining option is to leverage synthetic data. This is potentially feasible by leveraging simulators, or digital twins, of real environments \cite{bauer2024comprehensive,pezoulas2024synthetic,ruah2025bridge} or by adopting powerful general-purpose models such as LLMs \cite{park2025adaptivepredictionpowered}. As illustrated in Figure \ref{fig:1}(d), synthetic data can be used both to improve training and to enhance the calibration of a pre-trained model. In this context, we review prediction-powered inference, which combines small labeled samples with abundant synthetic predictions for training \cite{angelopoulos2023ppi,DRjordan}, as well as synthetic-powered predictive inference, which integrates synthetic data for calibration \cite{bashari2025synthetic,bashari2025statistical}.
\end{itemize}

The remainder of this paper is structured as follows. Section \ref{sec:II} introduces Bayesian learning and post-Bayesian variants as principled tools to quantify epistemic uncertainty. Section \ref{sec:III} presents information-theoretic generalization bounds that formalize the trade-off between training data and average accuracy. Section \ref{sec:IV} surveys predictive methods for uncertainty quantification based on conformal prediction and variants. Section \ref{sec:V} examines data efficiency via synthetic data through prediction-powered and synthetic-powered inference. Section \ref{sec:VI} provides concluding remarks and discusses future research directions. 

\begin{figure*}[!t]
\centering
 \includegraphics[width=0.7\linewidth]{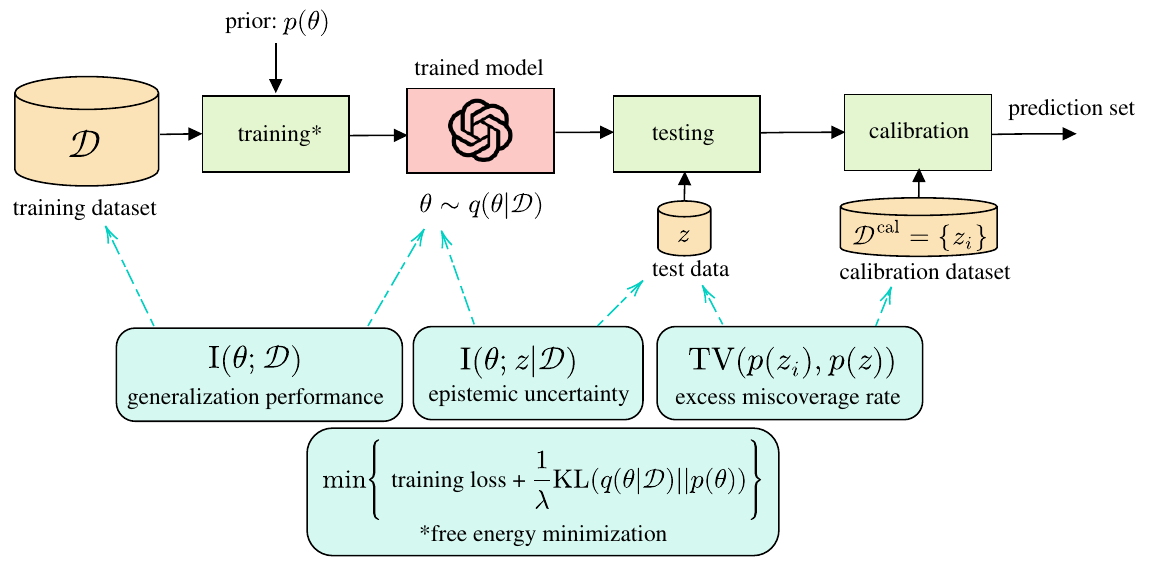}
\caption{Summary of information-theoretic relationships reviewed in this paper.}\label{fig:2}
\end{figure*}

The definition of the information-theoretic measures used in this article is provided in Table \ref{tab:information_measures}, while Table \ref{tab:acronyms} reports a table of acronyms. A summary of information-theoretic relationships surveyed in this paper can be found in Figure \ref{fig:2}.

\begin{table}[htbp]
\centering
\caption{Information-theoretic measures and their definitions.}
\label{tab:information_measures}
\begin{tabular}{ll}
\toprule
Measure & Definition\\
\midrule
Entropy & $\mathrm{H}(a)=-\mathbb{E}_{a\sim p(a)}\left[\log(p(a))\right]$\\
Conditional entropy & $\mathrm{H}(a|b)=-\mathbb{E}_{a,b\sim p(a,b)}\left[\log(p(a|b))\right]$\\
Mutual information & $\mathrm{I}(a;b)=\mathrm{H}(a)-\mathrm{H}(a|b)$\\
KL divergence & $\mathrm{KL}(p(a)\|q(a))=\mathbb{E}_{a\sim p(a)}\left[\log\left(\frac{p(a)}{q(a)}\right)\right]$\\
Total variation & $\mathrm{TV}(p(a),q(a))=\frac{1}{2}\|p-q\|_{1}$\\
\bottomrule
\end{tabular}
\end{table}

\section{Generalized Bayesian Learning}\label{sec:II}

Classical Bayesian inference provides a principled framework for incorporating prior knowledge and quantifying epistemic uncertainty through posterior distributions over parameters affecting the data distribution. The principle can be directly applied to the training of AI models, in which a posterior distribution over the model parameters (denoted as $\theta$) is inferred by combining prior information with a likelihood function \cite{theodoridis2015machine}. As shown in Figure \ref{fig:1}(a), this leads to a distribution in the model parameter space, whose extent and shape capture the uncertainty arising from the availability of limited data. In practice, the posterior distribution is approximated using sampling, variational methods, or Laplace approximations \cite{angelino2016patterns,simeone2022machine}.

\begin{table}[htbp]
\centering
\caption{Acronyms used in this paper.}
\label{tab:acronyms}
\begin{tabular}{ll}
\toprule
Acronym & Definition\\
\midrule
AI & Artificial intelligence\\
LLM & Large language model\\
CP & Conformal prediction\\
CRC & Conformal risk control\\
ERM & Empirical risk minimization\\
MCMC & Markov chain Monte Carlo\\
PAC & Probably approximately correct\\
PFN & Prior-fitted network\\
PPI & Prediction-powered inference\\
RCPS & Risk-controlling prediction set\\
RLHF & Reinforcement learning from human feedback \\
SPI & Synthetic-powered predictive inference\\

\bottomrule
\end{tabular}
\end{table}

The validity of the posterior distribution hinges on the adherence of prior and likelihood to the ground-truth data distribution. However, prior distributions are often chosen for their analytical tractability, and typical likelihood models fail to capture aspects such as outliers \cite{knoblauch2022optimization,zecchinPAC}. To obviate these issues, so-called ``post-Bayes'' methods have been introduced. These move beyond conventional Bayesian inference, while retaining the key goal of quantifying epistemic uncertainty. After reviewing conventional Bayesian learning, this section provides a brief introduction to generalized Bayesian learning,  martingale posteriors, and prior-fitted networks as  important representative post-Bayes frameworks.

\subsection{Bayesian Learning}

 Bayesian inference of  parameters $\theta$ begins by 
specifying a prior distribution $p(\theta)$ and a likelihood function 
$p(z|\theta)$, where $z$ denotes a data point. { The prior 
$p(\theta)$ should ideally encode available knowledge about the structure 
and typical values of parameters $\theta$ for which the likelihood 
$p(z|\theta)$ closely approximates the true data-generating distribution.} 
In supervised learning, the likelihood $p(z|\theta)$ takes the form of a 
conditional distribution $p(y|x,\theta)$, where $x$ is the input and $y$ 
is the output variable. { More broadly, the likelihood 
$p(z|\theta)$ defines a model class that encodes the learner's inductive 
bias \cite{shalev2014understanding}. As we discuss further in 
Section~\ref{sec:genBayeslearn}, it can also be derived from the loss 
function of a machine learning algorithm.}

Given a training {dataset (i.e., a multiset with possible multiplicities)}  $\mathcal{D} = \{z_1, \ldots, z_n\}$,  the posterior distribution is given by
\begin{equation}\label{eq:post}
p(\theta|\mathcal{D}) \propto {p(\theta) \prod_{i=1}^n p(z_i|\theta)},
\end{equation}
where the proportionality factor is evaluated to ensure normalization. 

Unlike conventional frequentist learning, which returns a single model parameter $\theta$, the posterior distribution (\ref{eq:post}) directly quantifies epistemic uncertainty through the spread of probability mass in the parameter space. Wide posterior distributions indicate high uncertainty about which parameters best explain the data, signaling insufficient training samples. As the training dataset grows in size, i.e., in the limit $n \to \infty$, consistency results show that, under appropriate conditions, the  posterior concentrates around the optimal parameter $\theta^*$, as illustrated in Figure \ref{fig:1}(a). Specifically, the optimal parameter $\theta^*$  provides the best fit for the true data distribution in terms of the Kullback-Leibler (KL) divergence (see Table \ref{tab:information_measures})  \cite{miller2021asymptotic}. 

Trained predictive or generative models can be obtained by drawing one or more samples from the distribution $p(\theta|\mathcal{D})$, typically using Markov chain Monte Carlo (MCMC) methods \cite{angelino2016patterns}. For example, the unadjusted Langevin algorithm produces the sequence of samples \begin{equation}
\theta_{k+1} = \theta_k + \eta \nabla_\theta \log p(\theta_k|\mathcal{D}) + \sqrt{2\eta} \, \xi_k, \quad \xi_k \sim \mathcal{N}(0, I),
\label{eq:ula}
\end{equation}
where $\xi_k$ are independent standard Gaussian random vectors and $\eta$ is the learning rate. Iterating the update \eqref{eq:ula} produces samples that approximate draws from the posterior $p(\theta|\mathcal{D})$ as $k \to \infty$ and $\eta \to 0$ \cite{welling2011bayesian}. Note that the update (\ref{eq:ula}) corresponds to a noisy version of standard gradient descent, and that there exist generalizations based on stochastic gradient descent \cite{welling2011bayesian} (see also \cite{simeone2022machine}).

With $K$ independent  samples $\theta_k\sim p(\theta|\mathcal{D})$ for $k=1,\ldots,K$, one can construct the \textbf{mixture} model
\begin{equation}\label{eq:ens}
q(z|\mathcal{D})=\frac{1}{K}\sum_{k=1}^K p(z|\theta_k), 
\end{equation}which is constructed from the \textbf{ensemble} of models $\{p(z|\theta_k)\}_{k=1}^{K}$. 
When the number of samples, $K$, is sufficiently large, the mixture distribution tends to the true data distribution given the training dataset $\mathcal{D}$,  \begin{equation}\label{eq:ens1}
p(z|\mathcal{D})=\mathbb{E}_{\theta\sim p(\theta|\mathcal{D})}[p(z|\theta)],
\end{equation}which is obtained by marginalizing over the posterior distribution $p(\theta|\mathcal{D})$.

Epistemic uncertainty is captured by the \textbf{diversity} of the ensemble of predictive distributions $\{p(z|\theta_k)\}_{k=1}^K$. In particular, in the case of supervised learning, given an input $x$, the epistemic uncertainty can be quantified by the degree to which the distributions  in the ensemble  $\{p(y|x,\theta_k)\}_{k=1}^K$  differ from one another.

The diversity of the ensemble, and thus the level of epistemic uncertainty, can be captured by metrics such as the variance of the top-1 predictive distributions $\{\max_yp(y|x,\theta_k)\}_{k=1}^K$ or by information-theoretic measures. A notable such measure is obtained by considering the difference between the entropy of the  predictive distribution (\ref{eq:ens1}) and the average entropy across the ensemble members, i.e., \cite{houlsby2011bayesian,simeone2022machine}
\begin{equation}\label{eq:BALD}
\mathrm{I}(\theta;z|\mathcal{D})= \mathrm{H}(z|\mathcal{D})-\mathbb{E}_{\theta \sim p(\theta|\mathcal{D})}[\mathrm{H}(z|\theta)],
\end{equation}
where $\mathrm{H}(z|\mathcal{D})$ is the entropy of the random variable $z \sim p(z|\mathcal{D})$ and $\mathrm{H}(z|\theta)$ is the entropy of the random variable $z \sim p(z|\theta)$. It is emphasized that (\ref{eq:BALD}) abuses the conditional entropy notation in Table \ref{tab:information_measures}, since the dataset $\mathcal{D}$ and the parameter $\theta$ are fixed (not random) in the entropies $\mathrm{H}(z|\mathcal{D})$ and $\mathrm{H}(z|\theta)$, respectively.

The quantity (\ref{eq:BALD}) is a measure of the diversity of the distributions $\{p(z|\theta)\}$ obtained by sampling the model parameters as $\theta \sim p(\theta|\mathcal{D})$. In fact, the mutual information in (\ref{eq:BALD})  equals zero if the distribution $p(z|\theta)$ does not depend on $\theta$, and thus the members of any ensemble  $\{p(z|\theta_k)\}_{k=1}^K$ are minimally diverse, all \textbf{agreeing} with each other. Conversely,  the measure (\ref{eq:BALD}) is large if the distributions $p(z|\theta)$ tend to \textbf{disagree} with each other for different samples $\theta \sim p(\theta|\mathcal{D})$, yielding  more diverse ensembles.  Note that  the mutual information in (\ref{eq:BALD}) is computed under the assumed prior $p(\theta)$ and likelihood $p(z|\theta)$. Thus, in order for the conditional mutual information in (\ref{eq:BALD}) to be a meaningful measure of epistemic uncertainty, one must assume prior and likelihood to be well-specified.  


 Information-theoretically, as indicated by the notation in (\ref{eq:BALD}), the difference between the two entropies corresponds to the mutual information $\mathrm{I}(\theta;z|\mathcal{D})$ between the model parameter and the observation $z$ conditioned on a fixed dataset $\mathcal{D}$ (see Figure \ref{fig:2}). This is a measure of the amount of information that we can obtain about the true model parameter $\theta$ by observing another data point $z$,  given that we already have access to the dataset $\mathcal{D}$. 
 
 { 
This epistemic uncertainty measure, given by the mutual information (\ref{eq:BALD}) between 
parameters and data, plays a central role in \textbf{Bayesian active learning}, where 
it is used to select the most informative data points for labeling 
\cite{houlsby2011bayesian}. Beyond active 
learning, mutual information-based epistemic uncertainty estimates have been 
leveraged, among other applications, for out-of-distribution detection \cite{malinin2018predictive} and safe 
decision-making in reinforcement learning \cite{depeweg2018decomposition}.}

\subsection{Generalized Bayesian Learning}\label{sec:genBayeslearn}

In practice, neither prior nor likelihood are typically guaranteed to reflect the true data-generating mechanism. To improve robustness to such modeling errors, the generalized Bayesian approach supports  more flexibility in the specification of the likelihood, as well as in the reliance of the (generalized) posterior distribution on the prior distribution.

To start, the generalized Bayesian methodology replaces the likelihood with a loss-based measure of the fit of model $\theta$ for data $z$. Let $\ell(\theta, z)$ denote such a loss function evaluating parameter $\theta$ on data point $z$. The loss function may, for instance, be the log-loss $\ell(\theta, z)=-\log p(z|\theta)$, or a generalization thereof that is more robust to the presence of outliers such as the $\alpha$-log-loss \cite{sypherd2022tunable,zecchinPAC,simeone2024classical}. The \textbf{training loss} is defined as the empirical average of the loss over the training data
\begin{equation}\label{eq:train}
L_{\text{train}}(\theta,\mathcal{D}) = \frac{1}{n}\sum_{i=1}^n \ell(\theta, z_i).
\end{equation}

The \textbf{generalized posterior}, or \textbf{Gibbs posterior}, is defined as
\begin{equation}\label{eq:genpost}
q(\theta|\mathcal{D}) \propto {p(\theta) \exp\left(-\lambda \cdot n \cdot L_{\text{train}}(\theta,\mathcal{D})\right)},
\end{equation}
where the hyperparameter $\lambda > 0$ controls the influence of the data relative to the prior. When the loss $\ell(\theta, z)$ equals the log-loss and $\lambda=1$, the Gibbs posterior reduces to the standard Bayesian posterior (\ref{eq:post}), establishing generalized Bayesian learning as a strict extension of conventional Bayesian learning.

Generalized Bayesian learning can be further extended so as to control the reliance of the generalized posterior distribution on the prior distribution. This is done by  observing that the generalized posterior $q(\theta|\mathcal{D})$ in (\ref{eq:genpost}) is the minimizer of the following information-theoretic objective, known as \textbf{free energy} \cite{knoblauch2022optimization,simeone2022machine,perlaza2024empirical}
\begin{equation}\label{eq:free}
 n\mathbb{E}_{\theta \sim q(\theta|\mathcal{D})}\left[L_{\text{train}}(\theta,\mathcal{D})\right] + \frac{1}{\lambda} \text{KL}\left(q(\theta|\mathcal{D}) \| p(\theta)\right),
\end{equation}
where $\text{KL}\left(\cdot \| \cdot\right)$ represents the KL divergence (see Table \ref{tab:information_measures}). The free energy objective (\ref{eq:free})  reveals that the generalized posterior distribution arises as a trade-off between fitting the training data, as required by the minimization of the first term in (\ref{eq:free}), and staying close to the prior, as enforced by the second term in (\ref{eq:free}) (see Figure \ref{fig:2}).

Replacing the KL divergence with other information-theoretic divergences, such as the R\'enyi and Tsallis relative entropies \cite{simeone2022machine}, yields a generalized form of the free energy objective (\ref{eq:free}). The minimizer of this objective represents an  extension of the notion of generalized posterior, which has been found to be useful in settings with a misspecified prior \cite{knoblauch2022optimization,zecchinPAC,daunas2026empirical}.

{
The  free energy (\ref{eq:free}) provides a principled framework for understanding 
regularized objectives in \textbf{LLM alignment}. Notably, in the context 
of \textbf{reinforcement learning from human feedback} (RLHF), we interpret $q$ as the LLM being optimized   
being optimized;  $p$ as a reference policy  corresponding to the pre-trained LLM; and the training loss $L_{\text{train}}$ as the negative reward  assigned by a 
learned reward model. The KL term thus penalizes deviation from the reference distribution, 
preventing the policy from exploiting spurious correlations in the imperfect reward signal,  
while preserving the general capabilities acquired during pre-training. The trade-off between maximizing expected reward 
(the first term) and remaining close to a trusted prior policy (the second term) is dictated by the parameter $\lambda$, whose selection is central to  RLHF-based alignment methods \cite{ouyang2022training}.}

\subsection{Martingale Posterior}

As discussed,  generalized Bayesian methods require the specification of a prior and of a likelihood function.  However, despite the added flexibility of generalized Bayesian techniques, it may be difficult to identify suitable choices on the basis of the available information about the problem. The \textbf{martingale posterior} approach  \cite{fong2023martingale} obviates this issue by requiring only the specification of a \textbf{predictive distribution},  allowing the generation of additional data given the available training dataset. Since there exist powerful foundation models covering a wide variety of data distributions of interest \cite{feuer2024tunetables}, the choice of a suitable predictive model may be practically less problematic than eliciting a prior and a likelihood.

More formally, the martingale posterior reframes posterior uncertainty about parameters as predictive uncertainty on \textbf{unseen, hypothetical data} conditional on the observed data. To model uncertainty over model parameters $\theta$, the martingale posterior specifies a predictive distribution $p(\mathcal{D}'|\mathcal{D})$ over unseen data $\mathcal{D}' = \{z_{n+1},\ldots,z_{n+n'}\}$ given the training data $\mathcal{D}=\{z_1,\ldots,z_n\}$, where $n'$ is an integer. In practice, the joint distribution $p(\mathcal{D}'|\mathcal{D})$ may be specified via one-step conditional distributions via the chain rule \begin{equation} p(\mathcal{D}'|\mathcal{D})=\prod_{i=1}^{n'} p(z_{n+i}|\mathcal{D},\{z_{n+j}\}_{j=1}^{i-1}).\end{equation}

Specifically, in the martingale posterior approach, one draws samples $\mathcal{D}' \sim p(\mathcal{D}'|\mathcal{D})$ of unseen data, and then solves the empirical risk minimization (ERM) problem
\begin{align}
\theta^{\text{MP}} &= \arg\min_{\theta} \sum_{z \in \mathcal{D} \cup \mathcal{D}'} \ell(z, \theta)\nonumber\\
&=\arg\min_{\theta} \left\{nL_{\text{train}}(\theta,\mathcal{D})+ \sum_{z \in \mathcal{D}'} \ell(z, \theta)\right\}
\label{eq:mp_erm}
\end{align}
over the combined dataset $\mathcal{D} \cup \mathcal{D}'$ encompassing both the training dataset $\mathcal{D}$ and the fictitious data $\mathcal{D}'$. The parameters $\theta^{\text{MP}}$ in \eqref{eq:mp_erm} are random variables due to the stochasticity of the unseen data $\mathcal{D}' \sim p(\mathcal{D}'|\mathcal{D})$. Samples $\theta^{\text{MP}}$ obtained by drawing independent realizations of unseen data $\mathcal{D}'$ are treated as draws from the underlying implicit {martingale posterior}.

The connection between samples from the conventional posterior \eqref{eq:post} and from the martingale posterior underlying the sequence of samples $\theta^{\text{MP}}$ in \eqref{eq:mp_erm} is given by De Finetti's theorem. Accordingly, if the predictive distribution $p(\mathcal{D}'|\mathcal{D})$ is exchangeable given $\mathcal{D}$ (i.e., if the distribution does not depend on the ordering of the samples $\mathcal{D}'$), the unseen samples $\mathcal{D}'$ can be conceptually thought of as being generated via the following ancestral sampling scheme:
\begin{enumerate}
\item Draw a sample $\theta$ from the posterior distribution $p(\theta|\mathcal{D})$;
\item Draw the unseen data $\mathcal{D}' \sim p(\mathcal{D}'|\theta) = \prod_{i=1}^{n'} p(z_{n+i}|\theta)$.
\end{enumerate}
In this case, thanks to the consistency of the maximum likelihood estimator, choosing the log-loss $\ell(z, \theta) = -\log p(z|\theta)$ in \eqref{eq:mp_erm} ensures that the distribution of the samples $\theta^{\text{MP}}$ converges in distribution to the posterior $p(\theta|\mathcal{D})$ as $n' \to \infty$ under weak regularity conditions \cite{fong2023martingale}.

{
\subsection{Prior-Fitted Networks}
\label{sec:pfn}

As discussed in this section, a central challenge in deploying Bayesian learning in practice is the 
computational cost of posterior inference. Standard methods, such as MCMC sampling methods or variational inference \cite{angelino2016patterns,simeone2022machine},  require iterative 
optimization or sampling procedures, such as (\ref{eq:ula}), that must be repeated from scratch for 
every new dataset. This difficulty is compounded 
when the prior encodes complex, domain-specific generative assumptions, e.g., priors defined through structural causal models or  simulators. 

\textbf{Prior-fitted networks} (PFNs) \cite{muller2021transformers}  recasts Bayesian inference as a supervised 
meta-learning problem, leveraging the \textbf{in-context learning} (ICL) capabilities of modern sequence models. Specifically, a transformer network is trained on a large 
collection of synthetic datasets, each drawn from the given prior. Given a 
training dataset of input-output pairs as context, the network learns to 
predict the posterior predictive distribution for held-out query points 
in a single forward pass without any iterative adaptation at test 
time. This way,  the cost of 
learning is shifted from test time to a one-time pretraining phase, amortizing the cost across an arbitrary number of Bayesian inference tasks.

Applications of PFNs include Bayesian optimization \cite{muller2023pfns4bo}, tabular models \cite{hollmann2022tabpfn}, and  causal inference 
\cite{balazadeh2025causalpfn}. Further discussions on the use of synthetic data can be found in Section \ref{sec:V}.

}

\section{Information-Theoretic Generalization Bounds}\label{sec:III}

As illustrated in Figure \ref{fig:1}(b), generalization bounds quantify the gap between population loss -- the metric ideally minimized during the training of an AI model -- and the training loss -- the metric actually minimized based on the available data. These bounds provide insights into the training data requirements, potentially offering guidance for algorithm design and data collection strategies. This section briefly reviews generalization bounds that leverage information-theoretic metrics. It will be shown that such bounds provide a  theoretical justification for the use of generalized Bayesian learning.

\subsection{Bounding the Generalization Error}

Given a model class, e.g., a class of neural networks, consider a learning algorithm that produces model parameters $\theta$ from the training dataset $\mathcal{D} = \{z_1, \ldots, z_n\}$, where each data point $z_i = (x_i, y_i)$ is drawn i.i.d. from a given distribution $p(z)$. The trained model parameters $\theta$ are typically random functions of the training dataset $\mathcal{D}$. In fact, the output of standard training algorithms such as stochastic gradient descent depends on factors such as random initialization and random mini-batch selection. Furthermore, the generalized Bayesian strategies reviewed in the previous section are often implemented to produce samples $\theta \sim q(\theta|\mathcal{D})$ from the generalized posterior, and the martingale posterior, also discussed in Section \ref{sec:II}, generates samples $\theta$ via (\ref{eq:mp_erm}).

We denote the conditional distribution of the model parameters given the training data as $q(\theta|\mathcal{D})$, so that the trained model is generated as
\begin{equation}
\theta \sim q(\theta|\mathcal{D}).
\end{equation}
The distribution $q(\theta|\mathcal{D})$ can be, for instance, the generalized posterior (\ref{eq:genpost}) or the distribution of the martingale posterior samples (\ref{eq:mp_erm}), but no such restrictions are imposed in this subsection.

To define the generalization error of a training algorithm, with reference to Figure \ref{fig:1}(b), we define the \textbf{population loss}, also known as \textbf{risk},  as the expected value
\begin{equation}
L_{\text{pop}}(\theta) = \mathbb{E}_{z\sim p(z)}[\ell(\theta, z)].
\end{equation}
The population loss represents the ideal target to be minimized via training, and the training loss $L_{\text{train}}(\theta,\mathcal{D})$ in \eqref{eq:train} is an unbiased estimate of the population loss. The \textbf{generalization error} is then given by the difference
\begin{equation}
\text{gen}(\theta,\mathcal{D}) = L_{\text{pop}}(\theta) - L_{\text{train}}(\theta,\mathcal{D}).
\end{equation}
We are interested in the generalization error $\text{gen}(\theta,\mathcal{D})$ of the trained model $\theta\sim q(\theta|\mathcal{D})$.

Intuitively, as illustrated in Figure \ref{fig:1}(b), the generalization error  for model $\theta\sim q(\theta|\mathcal{D})$ should decrease as the dataset $\mathcal{D}$ grows in size, since the training loss becomes a more accurate estimate of the population loss. In particular, generalization is known to be problematic when the trained model parameters $\theta$ exhibit an excessive dependence on the specific realization of the training dataset $\mathcal{D}$ -- a phenomenon known as \textbf{overfitting}.

{\subsection{Bounding the Average Generalization Error}}
Validating this observation, seminal work \cite{xu2017information} has shown that, for loss functions satisfying a $\sigma$-sub-Gaussian tail condition, the expected value of the generalization error can be upper bounded as
\begin{equation}
\mathbb{E}_{\mathcal{D},\theta \sim p(\mathcal{D})q(\theta|\mathcal{D})}[\text{gen}(\theta,\mathcal{D})] \leq \sqrt{\frac{2\sigma^2 \mathrm{I}(\mathcal{D}; \theta)}{n}},
\label{eq:mi_bound}
\end{equation}
where $\mathrm{I}(\mathcal{D}; \theta)$ represents the mutual information of the random variables $\mathcal{D},\theta \sim p(\mathcal{D})q(\theta|\mathcal{D})$. Note that the expected value in (\ref{eq:mi_bound}) is also evaluated over the joint distribution of the training data $\mathcal{D}$ and of the output of the training algorithm $\theta \sim q(\theta|\mathcal{D})$. Accordingly, the mutual information $\mathrm{I}(\mathcal{D}; \theta)$ between the training dataset and the learned hypothesis quantifies how much information about the random training data is captured by the algorithm's output (see Figure \ref{fig:2}). Lower mutual information indicates the algorithm produces more stable outputs across different training sets, suggesting better generalization. Extensions of this bound are reviewed in \cite{hellstrom2023generalization}.

For specific algorithm classes, such as stochastic gradient descent with appropriate step sizes, the mutual information $\mathrm{I}(\mathcal{D}; \theta)$ grows sub-linearly with $n$, e.g., as $O(\log n)$ \cite{hellstrom2023generalization}. This ensures the right-hand side of the bound \eqref{eq:mi_bound} vanishes as $n \to \infty$, providing theoretical justification for the effectiveness of the corresponding training algorithms in practice. Moreover, the information-theoretic bound (\ref{eq:mi_bound}) may also guide algorithm design: techniques like gradient noise injection, dropout, and data augmentation can be understood as mechanisms to reduce the mutual information $\mathrm{I}(\mathcal{D}; \theta)$, thereby reducing the risk of overfitting and hence the generalization error. { Finally, exact expressions for the average generalization error have been derived for the Gibbs posterior (\ref{eq:genpost}) \cite{aminian2023information}.}

{\subsection{Bounding the  Generalization Error with High Probability}}

It is also possible to derive \textbf{high-probability} upper bounds on the population loss as a function of the training loss, offering another way to bound the generalization error. Specifically, the \textbf{probably approximately correct (PAC)-Bayes} framework yields bounds of the following form. Fix any constant $\lambda>0$; assume that the loss takes values in the interval $[0,C]$; and choose  any prior distribution $p(\theta)$ on the model parameter space, as long as it is  independent of the training data. Then, for any $\delta \in (0,1)$, with probability at least $1-\delta$ over the random draw of the training dataset $\mathcal{D}$ of size $n$, the following bound holds \cite{catoni2007pac,alquier2024user}
\begin{equation}\label{eq:PAC}
\mathbb{E}_{\theta \sim q({\theta}|\mathcal{D})}[\text{gen}(\theta,\mathcal{D})] \leq \frac{\lambda C^2}{8n}+{\frac{\text{KL}(q({\theta}|\mathcal{D})\|p(\theta))+\log\left(\frac{1}{\delta}\right)}{\lambda}}.
\end{equation}

The PAC-Bayes bound \eqref{eq:PAC} quantifies the generalization error via the KL divergence term $\text{KL}(q({\theta}|\mathcal{D}) \| p(\theta))$, which provides another measure of the dependence of the trained model on the training data $\mathcal{D}$. With a suitable choice of the prior $p(\theta)$ (namely, $p(\theta)=\mathbb{E}_{\mathcal{D}\sim p(\mathcal{D})}[q(\theta|\mathcal{D})]$), the average of this term over the training data distribution indeed coincides with the mutual information in (\ref{eq:mi_bound}) (i.e., $\mathbb{E}_{\mathcal{D}\sim p(\mathcal{D})}[\text{KL}(q({\theta}|\mathcal{D}) \| p(\theta))]=\mathrm{I}(\mathcal{D}; \theta)$) \cite{hellstrom2023generalization}.

{ PAC-Bayes bounds have 
emerged as one of the most effective theoretical tools for studying 
generalization in modern deep learning. For example, references \cite{dziugaite2017computing,lotfi2023non}  demonstrated  
non-vacuous generalization bounds for deep neural networks and LLMs, respectively. These 
results support the broader thesis, articulated by 
\cite{wilson2025deep}, that phenomena such as benign overfitting, double 
descent,  can be rigorously characterized by 
long-standing frameworks like PAC-Bayes that reward flexible hypothesis 
spaces endowed with soft inductive biases favoring simpler, more 
compressible solutions.}

The inequality \eqref{eq:PAC} can be rewritten as an upper bound on the population loss,  averaged over the output of the training algorithm, as \begin{align}
\mathbb{E}_{\theta\sim q({\theta}|\mathcal{D})}[L_{\text{pop}}(\theta)] & \leq  \mathbb{E}_{\theta\sim q({\theta}|\mathcal{D})}[L_{\text{train}}(\theta,\mathcal{D})]\nonumber\\
 & +\frac{\lambda C^2}{8n}+{\frac{\text{KL}(q({\theta}|\mathcal{D})\|p(\theta))+\log\left(\frac{1}{\delta}\right)}{\lambda}}. \label{eq:PAC1}
\end{align} Since the bound \eqref{eq:PAC}, and thus also \eqref{eq:PAC1}, hold simultaneously for all distributions $q({\theta}|\mathcal{D})$, the upper bound \eqref{eq:PAC1} can be optimized over the distributions $q({\theta}|\mathcal{D})$. This yields a training objective that is closely related to the free energy objective (\ref{eq:free}). In fact, both (\ref{eq:free}) and (\ref{eq:PAC1}) are minimized by the generalized posterior (\ref{eq:genpost}). This shows that the generalized posterior distribution is the minimizer of an upper bound on the population loss, justifying its use in settings with limited training data.

\section{Uncertainty Quantification}\label{sec:IV}

While generalized Bayesian learning provides a useful framework for understanding epistemic uncertainty with asymptotic validity properties in the model parameter space, practical deployments often require methods that have finite-sample guarantees on the predictive performance in the output space. This section reviews conformal prediction and related techniques that achieve distribution-free uncertainty quantification with minimal assumptions.

\subsection{Set Prediction}

Set-valued predictions provide a convenient and actionable way to quantify uncertainty in supervised learning settings. Specifically, given an input $x$, a \textbf{set-valued predictor} $\mathcal{C}(x)$ maps the input to a subset of the output space. For example, in Figure~\ref{fig:1}(c), the set $\mathcal{C}(x)$ represents the confidence region (error bars) around the trained predictor, shown as the red shaded area. For a desired miscoverage rate $\alpha \in (0,1)$, e.g., $10\%$ or $\alpha=0.1$,  a useful property for the set predictor is to satisfy the \textbf{coverage guarantee}
\begin{equation}\label{eq:coverage}
\mathbb{P}(y \in \mathcal{C}(x)) \geq 1 - \alpha,
\end{equation}
where the probability is evaluated over the joint distribution of input $x$ and output $y$. Accordingly, the true output $y$ is contained within the set $\mathcal{C}(x)$ with probability no smaller than $1-\alpha$. 

When this condition is satisfied, the prediction set $\mathcal{C}(x)$ provides quantifiable information about the uncertainty associated with the model's output. Such insights may enable reliable decision-making in downstream applications. For instance, safety-critical control protocols can ensure that safety constraints hold for any value of the target $y$ within the set $\mathcal{C}(x)$, thereby meeting reliability requirements with probability at least $1-\alpha$ \cite{lindemann2023safe,kiyani2025decision,zecchin2024forking}.

If the ground-truth conditional distribution $p(y|x)$ of the target $y$ given input $x$ were known, the smallest set $\mathcal{C}(x)$ satisfying \eqref{eq:coverage} could be directly obtained as the \textbf{highest-probability region} under the distribution $p(y|x)$. This optimal set can be expressed as
\begin{equation} 
C^*(x) = \left\{y: p(y|x) \geq \tau(x)\right\},
\end{equation}
where the threshold $\tau(x)$ is chosen as the smallest value such that the condition 
\begin{equation}
\sum_{y \in C^*(x)} p(y|x) \geq 1-\alpha
\end{equation}
holds, or equivalently,
\begin{equation}\label{eq:ideal}
C^*(x) = \left\{y: \sum_{y':p(y'|x)\geq p(y|x)} p(y'|x) \leq 1-\alpha\right\}.
\end{equation}
For discrete output variables $y$, this set can be constructed by listing the probabilities $\{p(y|x)\}_y$ in non-increasing order for the given input $x$, and then including in set $C^*(x)$ the values of $y$ corresponding to the top elements of this ordered list until the cumulative probability reaches or exceeds the level $1-\alpha$.

Importantly, when constructed using the true distribution $p(y|x)$, the ideal set $C^*(x)$ satisfies the stronger \textbf{conditional coverage} property
\begin{equation}\label{eq:conditional_coverage}
\mathbb{P}(y \in C^*(x)|x) = 1 - \alpha,
\end{equation}
which holds for each fixed value of $x$. This conditional guarantee is more informative than the marginal coverage in \eqref{eq:coverage}, as it ensures the specified coverage level for every individual input $x$ rather than only on average over the input distribution.

The \textbf{average} size of the optimal prediction set \eqref{eq:ideal} can be related to the uncertainty in the conditional distribution $p(y|x)$. Specifically, the average size of the set $C^*(x)$ can be bounded in terms of the conditional entropy $\mathrm{H}(y|x)$ of the output $y$ given input $x$ \cite{correia2024information}.

Moreover, considering any fixed  input $x$, one expects that regions of the input space with a higher-entropy distribution $p(y|x)$  yield larger prediction sets. Writing, with some abuse of notation with respect to Table \ref{tab:information_measures}, as  $\mathrm{H}(y|x)$ the entropy of the random variable $y\sim p(y|x)$ for a fixed $x$, this observation can be formalized as follows. Given two inputs $x$ and $x'$, if the prediction set $C^*(x)$ is always larger than the corresponding set  $C^*(x')$ for all miscoverage levels $\alpha$,  the probability vector with entries $\{p(y|x)\}_y$ is \textbf{majorized} by the vector $\{p(y|x')\}_y$ \cite{marshall1979inequalities}.  This, in turn, implies the inequality $\mathrm{H}(y|x) \geq \mathrm{H}(y|x')$ between the entropies of the random variables $y \sim p(y|x)$ and $y' \sim p(y|x')$   \cite{marshall1979inequalities,simeone2024classical,wang2024credal}.

\subsection{Conformal Prediction}

\textbf{Conformal prediction (CP)} provides a framework for constructing prediction sets with guaranteed marginal coverage (\ref{eq:coverage}), regardless of the underlying data distribution \cite{vovk2005algorithmic,angelopoulos2024theoretical}. At first glance, ensuring the condition (\ref{eq:coverage}) appears to be an impossible goal without knowledge of the true conditional distribution $p(y|x)$.  In fact, any trained predictive distribution $p(y|x,\theta)$, even when extended to an ensemble as in (\ref{eq:ens}), generally does not match the true conditional probability $p(y|x)$. This discrepancy corresponds to the well-known problem of \textbf{miscalibration} of machine learning models \cite{guo2017calibration,huang2024calibrating,kapoor2024large}.  Formally, it can, in fact, be proved that it is impossible to ensure the \textbf{conditional} coverage inequality $\mathbb{P}(y \in \mathcal{C}(x)|x) \geq 1 - \alpha$, where the probability is conditioned on the input $x$, unless strong assumptions are made on the conditional distribution of the output given the input \cite{foygel2021limits}.

CP bypasses this fundamental limitation by weakening the guarantee to a marginal one:  evaluating the probability (\ref{eq:coverage}) not only over the distribution of the output $y$, but also over the input $x$ and over calibration data used to evaluate the set $\mathcal{C}(x)$. Therefore, from a frequentist perspective, the probability (\ref{eq:coverage}) reflects the fraction of realizations of inputs $x$, outputs $y$, and calibration datasets for which the target $y$ is found to lie in the prediction set $\mathcal{C}(x)$. This requirement can be alleviated in various ways by imposing different notions of partial locality, and the reader is referred to references \cite{vovk2012conditional,gibbs2023conditional,hore2023conformal,romano2022classification} for further discussion on this point.

The most computationally efficient variant is \textbf{split CP} \cite{angelopoulos2024theoretical}. The procedure partitions the available data  into training set $\mathcal{D}$ and calibration dataset $\mathcal{D}_{\text{cal}} = \{(x_i, y_i)\}_{i=1}^m$, and it applies the following steps:

\noindent\textbf{Step 1: Training.} Train a model $q(y|x)$ using training dataset $\mathcal{D}$. Note that the model may be trained using conventional frequentist learning or via Bayesian methods, producing an ensemble predictor as in (\ref{eq:ens}).

\noindent\textbf{Step 2: Evaluate non-conformity calibration scores.} Using the trained model $q(y|x)$, compute \textbf{non-conformity scores} $s = s(x, y)$ -- henceforth referred to as scores for short -- for each (labeled) calibration point $(x,y)\in\mathcal{D}_{\text{cal}}$. Intuitively, the scores represent the loss (or prediction error) of the pre-trained model $q(y|x)$ on each calibration data point. Evaluating statistics of the calibration scores thus provides insights into the typical behavior of the model. Some examples are provided below.

\noindent\textbf{Step 3: Prediction set construction via score analysis.} Given an input $x$, the prediction set $\mathcal{C}(x)$ is constructed by including all the values of $y$ whose score is sufficiently small. Specifically, the threshold is obtained by ordering the calibration scores and evaluating the $\lceil (1-\alpha)(m+1)\rceil$-th smallest score. This is denoted as
\begin{equation}\label{eq:quant}
\hat{Q}_\alpha = \text{Quantile}\left(\{s_1, \ldots, s_m\}, \lceil (1-\alpha)(m+1)\rceil/m\right),
\end{equation}
where  $s_i=s(x_i,y_i)$ is the score for the $i$-th calibration data point $(x_i,y_i)$. As indicated by the notation in (\ref{eq:quant}), this can be equivalently interpreted as the $\lceil (1-\alpha)(m+1)\rceil/m$-quantile of the empirical distribution of the calibration scores $\{s_1, \ldots, s_m\}$. Accordingly, for a new test point $x$, the CP set is evaluated as
\begin{equation}\label{eq:cpset}
\mathcal{C}(x) = \left\{y : s(x, y) \leq \hat{Q}_\alpha\right\}.
\end{equation}

\noindent \textbf{Common choices for the scores:} Common choices of score function include the following:
\begin{itemize}
\item \textbf{Loss for point predictors}: Given the model $q(y|x)$, obtain a point prediction $\hat{y}(x)$, e.g., as $\hat{y}(x)=\arg \max_y q(y|x)$, and then evaluate standard loss measures such as the squared loss $s(x,y) = (y - \hat{y}(x))^2$ for regression.

\item \textbf{Conformalized quantile regression score}: For scalar regression problems, using the pre-trained model (or models), obtain an estimated $\alpha^-$-quantile and an $\alpha^+$-quantile of the conditional distribution of the target $y$ given $x$, with probabilities $\alpha^+>\alpha^-$, which are denoted as $\hat{q}_{\alpha^-}(x)$ and $\hat{q}_{\alpha^+}(x)$, respectively. Then, evaluate the score as \cite{romano2019conformalized}
$$s(x,y) = \max\left\{ \hat{q}_{\alpha^-}(x) - y, \; y - \hat{q}_{\alpha^{+}}(x) \right\}.$$
 Typical choices for the probabilities defining the quantiles are   $\alpha^-=\alpha/2$ and  $\alpha^+=1-\alpha/2$.

\item \textbf{Loss for predictive distributions}: For classification, evaluate scores as functions of the probability $q(y|x)$, such as $s(x,y) = 1 - q(y|x)$ or the log-loss $s(x,y) = -\log (q(y|x))$.

\item \textbf{Adaptive prediction sets score}: Mimicking the ideal prediction set (\ref{eq:ideal}), the adaptive prediction set score lists the probabilities as $q(y|x)$ in non-increasing order, and then defines the score as the sum over all target values whose probability is higher than or equal to that for $y$, i.e., \cite{romano2020classification}
\begin{equation}\label{eq:APS} s(x, y) = \sum_{y': q(y'|x) \geq q(y|x)} q(y'|x).\end{equation} 

\item \textbf{Scores for multivariate target variables}: For multivariate target variables, there is a wide range of options that are tailored to the geometry of the target space, with tools ranging from latent-space analysis to optimal transport \cite{dheur2025unified,klein2025multivariate}.
\end{itemize}

In practice, the evaluation of the set (\ref{eq:cpset}) depends on the choice of the score. In particular, in some cases, there is an equivalent, computationally efficient formula for the prediction set in \eqref{eq:cpset} that does not require sweeping over all possible values of $y$.    For example, for the conformalized quantile regression score, we have
\begin{equation}
\hat{C}(x) = \left[\hat{q}_{\alpha^-}(x) - \hat{Q}_{\alpha}, \; \hat{q}_{\alpha^+}(x) + \hat{Q}_{\alpha}\right].
\end{equation}

\noindent\textbf{Theoretical guarantees:} The CP set (\ref{eq:cpset}) satisfies the desired coverage condition (\ref{eq:coverage}) as long as the examples in the calibration dataset $\mathcal{D}_{\text{cal}}$ and the test pair $(x,y)$ are exchangeable when viewed as a sequence of $m+1$ variables. For example, this is the case under the common assumption that calibration data and test data are i.i.d. Importantly, this result holds \textbf{distribution-free}, in the sense that no assumptions on the trained model and on data distribution (besides exchangeability) are required.

Consider now a more general situation in which each calibration data point $z_i=(x_i,y_i)$ has a generally different distribution $p(z_i)$, and denote as $p(z)$ the distribution of the test point $z=(x,y)$. We focus here on the special case of independent data, but the discussion can be extended. Assuming that each calibration data point $z_i$ is weighted by a coefficient $w_i$ in computing the quantile (\ref{eq:quant}), the coverage property (\ref{eq:coverage}) is modified as \cite{tibshirani2019conformal}
\begin{equation}\label{eq:coverage_shift}
\mathbb{P}(y \in \mathcal{C}(x)) \geq 1 - \alpha - 2\sum_{i=1}^{m} w_i \cdot {\mathrm{TV}}(p(z_i), p(z)),
\end{equation}
where ${\mathrm{TV}}(\cdot, \cdot)$ is the \textbf{total variation distance} between two distributions (see Table \ref{tab:information_measures}). Accordingly, a distribution shift, captured by the total variation distance ${\mathrm{TV}}(p(z_i), p(z))$, causes an increase in  the miscoverage rate beyond the target value $\alpha$ (see Figure \ref{fig:2}). This issue can be mitigated by assigning calibration data points $z_i$ with large expected data shifts a smaller weight $w_i$. For example, for a time series, it may be useful to give larger weights to more recent samples.

\subsection{Conformal Risk Control and Risk-Controlling Prediction Sets}

While standard CP controls the miscoverage rate as in (\ref{eq:coverage}), many applications require controlling the expected value of a more general loss function. \textbf{Conformal risk control (CRC)} \cite{angelopoulos2023crc} extends CP to control the expected risk $\mathbb{E}[\mathcal{L}(\mathcal{C}(x), y)]$, where $\mathcal{L}(\mathcal{C}, y)$ is a loss function that is non-increasing with the size of the prediction set $\mathcal{C}$. CP is recovered by choosing the loss $\mathcal{L}(\mathcal{C} y)={1}(y\notin \mathcal{C})$, where $1(\cdot)$ is the indicator function,  but other losses supported by CRC include false negative rates in segmentation and tracking \cite{angelopoulos2023crc,zecchin2024forking}, as well as ranking-dependent error measures \cite{xu2024two}.

 CRC ensures the inequality
\begin{equation}\label{eq:crc}
\mathbb{E}[\mathcal{L}(\mathcal{C}(x), y)] \leq \beta,
\end{equation}
where $\beta$ is a user-specified risk threshold. In a manner similar to (\ref{eq:coverage}), the expected value in \eqref{eq:crc} is evaluated with respect to the joint distribution of the test data $(x,y)$ and of the calibration data used to design the prediction set $\mathcal{C}(x)$.

The prediction set in CRC is obtained as in (\ref{eq:cpset}), with the caveat that the threshold $\hat{Q}_\alpha$ is chosen so as to ensure that the empirical risk $\sum_{i=1}^m \mathcal{L}(C(x_i), y_i)/m$ over the calibration data does not exceed $\beta$ plus a slack term dependent on the maximum value of the loss. The procedure generalizes CP.

In practice, it may be more relevant to ensure that the expected risk remains below the threshold $\beta$ with high probability with respect to the calibration dataset. That is, it may be preferable to ensure the condition
\begin{equation}\label{eq:rcps}
\mathbb{P}\{\mathbb{E}[\mathcal{L}(\mathcal{C}(x), y)] \leq \beta\}\geq 1-\delta,
\end{equation}
where the inner expected value is with respect to the test pair $(x,y)$, the outer probability is over the calibration data, and $\delta$ is a user-defined probability.

This goal can be obtained via \textbf{risk-controlling prediction sets (RCPS)} \cite{bates2021distribution}, which construct the prediction set by first obtaining an upper confidence bound, with coverage probability $1-\delta$, on the expected risk using the calibration data. Then, the threshold in the set (\ref{eq:cpset}) is evaluated by finding the smallest value of the threshold for which all larger values yield an upper bound no larger than the target $\beta$.

The validity of this approach rests on an interpretation of the construction of the prediction set as a form of \textbf{multiple hypothesis testing}. Accordingly, one views the choice of the threshold as the testing of multiple hypotheses, each corresponding to a different choice of the threshold. This principle can be generalized to the optimization of other hyperparameters \cite{angelopoulos2025learn,farzaneh2025ensuring}. In this regard, it is noted that CP can also be described from the perspective of hypothesis testing via the notion of conformal p-values \cite{angelopoulos2024theoretical}.

\subsection{On the Efficiency of Prediction Sets}

A fundamental question is: {How tight can prediction sets be while maintaining coverage?} Intuitively, the answer depends critically both on the generalization error of the model used to evaluate the score functions and on the amount of calibration data available to evaluate the threshold in (\ref{eq:quant}).

In \cite{zecchin2024generalization,zecchin2025generalization}, an information-theoretic upper bound on the expected size of CP and CRC sets is derived that takes the form
\begin{equation}
\mathbb{E}[|\mathcal{C}(x)|] \leq f\left(m, \alpha, \text{gen}(\theta,\mathcal{D})\right),
\end{equation}
where $m$ is the calibration dataset size, $\alpha$ is the target miscoverage rate, and $\text{gen}(\theta,\mathcal{D})$ is the generalization error of the pre-trained predictor used to evaluate the scores (see Section \ref{sec:III}). The function $f$ is increasing in the generalization error $\text{gen}(\theta,\mathcal{D})$ and decreasing in both $m$ and $\alpha$.

This bound shows that reducing generalization error directly improves prediction set informativeness. Furthermore, the bound exhibits an exponential decrease in the calibration dataset size, demonstrating that the demands in terms of calibration data are significantly more limited than in terms of training data. A typical choice is to ensure a number of calibration data points that is of the order $10/\alpha$--$100/\alpha$, growing inversely proportionally with the target miscoverage rate $\alpha$.

\section{Data Efficiency via Synthetic Data}\label{sec:V}

The previous sections have discussed ways to cope with the availability of limited data through uncertainty quantification. This section covers a complementary approach, whereby the dataset is augmented with auxiliary, synthetic information sources. Examples of this methodology include the use of simulators, or \textbf{digital twins}, of real-world systems to emulate the behavior of the true data distribution. This type of framework is increasingly adopted for practical deployments in the fields of robotics, telecommunications, and healthcare \cite{polese2024colosseum,ruah2025bridge}. Other instances of this approach encompass distillation \cite{woo2025synthetic} and weak-to-strong generalization \cite{burns2023weak}.

Synthetic data can be potentially used both to augment the training dataset, hence contributing to an improvement of the underlying AI model, and to increase the effective size of the calibration dataset, thus potentially yielding a more accurate quantification of uncertainty. In both cases, the main challenge in using synthetic data is the inherent bias caused by the use of data that does not follow precisely the real-world data distribution. This is also known as the \textbf{sim-to-real gap} when synthetic data are generated by simulators.

This section examines two recent frameworks that address the bias of synthetic data. The first, prediction-powered inference \cite{angelopoulos2023ppi,angelopoulos2024ppiplus,zrnic2024cross}, and the closely related doubly robust self-training approach \cite{DRjordan,sifaou2024semi,shohamprediction}, integrate synthetic data to enhance model training, while synthetic-powered predictive inference \cite{bashari2025synthetic,bashari2025statistical} addresses model calibration.

\begin{figure}[!t]
\centering
 \includegraphics[width=1.0\linewidth]{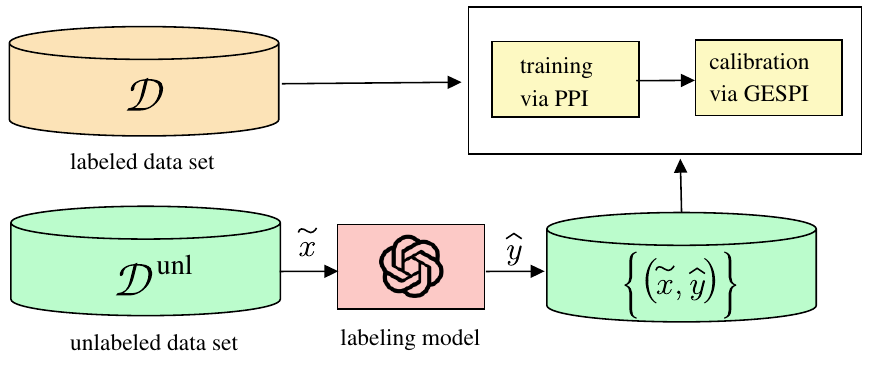}
\caption{Prediction-powered inference (PPI) \cite{angelopoulos2023ppi,angelopoulos2024ppiplus,zrnic2024cross}, and the closely related doubly robust self-training approach \cite{DRjordan,sifaou2024semi}, integrate synthetic data to enhance model training, while generalized synthetic-powered predictive inference (GESPI)  \cite{bashari2025synthetic,bashari2025statistical} addresses model calibration.}\label{fig:3}
\end{figure}

\subsection{Prediction-Powered Inference}

As illustrated in Figure \ref{fig:3}, \textbf{prediction-powered inference (PPI)} \cite{angelopoulos2023ppi,angelopoulos2024ppiplus,zrnic2024cross} assumes the availability of a small labeled dataset $\mathcal{D}=\{z_i=(x_i, y_i)\}_{i=1}^n$ with true outcomes $y_i$, and a much larger dataset $\mathcal{D}_{\mathrm{unl}}=\{(\tilde{x}_j)\}_{j=1}^N$ with inputs $\tilde{x}_j$ following the same distribution as the labeled dataset, where $N \gg n$. The goal is to estimate a vector $\theta$, such as the parameters of an AI model \cite{DRjordan,sifaou2024semi}, using both datasets to achieve better performance than using labeled data alone.

In PPI, we use an auxiliary labeling model to assign a synthetic label $\hat{y}_i$ to each labeled data point $(x_i,y_i)\in \mathcal{D}$, as well as a label $\hat{y}_j$ to each unlabeled data point $\tilde{x}_j\in \mathcal{D}_{\text{unl}}$. Focusing on model training \cite{DRjordan,sifaou2024semi}, the PPI estimator of the population loss integrates both real training data and synthetic data via the following linear combination:
\begin{align}\label{eq:PPI}
&L_{\mathrm{PPI}}(\theta,\mathcal{D},\mathcal{D}_{\text{unl}}) = \lambda\cdot\underbrace{\frac{1}{N}\sum_{j=1}^{N}\ell(\theta,(\tilde{x}_{j},\hat{y}_{j}))}_{L_{\text{train}}(\theta,\mathcal{D}_{\text{unl}}):\text{ training loss on dataset }\mathcal{D}_{\text{unl}}}\nonumber \\
&\quad +\underbrace{\left(\frac{1}{n}\sum_{i=1}^{n}\ell(\theta,(x_{i},y_{i}))-\lambda\cdot\frac{1}{n}\sum_{i=1}^{n}\ell(\theta,(x_{i},\hat{y}_{i}))\right)}_{\text{correction term using dataset }\mathcal{D}},
\end{align}
where $\lambda \geq 0$ is a hyperparameter that determines the extent to which the estimate (\ref{eq:PPI}) of the population loss relies on the synthetic labels. It is noted that the PPI estimator (\ref{eq:PPI}) relates to prior methods such as doubly robust estimators \cite{dudik2014doubly,breidt2017model}.

Intuitively, the correction term in (\ref{eq:PPI}) leverages the true labels to estimate the error caused by the use of the synthetic labels in lieu of the true labels. Note that evaluating this term is possible, since it is computed using labeled data. Thanks to the presence of this correction term, assuming the unlabeled covariates $\tilde{x}$ are drawn i.i.d. from the same distribution of the labeled covariates $x$,  the quantity \eqref{eq:PPI} can be readily seen to be an \textbf{unbiased} estimate of the population loss, i.e.,
\begin{equation}
\mathbb{E}[L_{\text{PPI}}(\theta,\mathcal{D},\mathcal{D}_{\text{unl}})]=L_{\mathrm{pop}}(\theta),
\end{equation}
where the average is over the labeled and unlabeled data, irrespective of the quality of the synthetic labels and for any value $\lambda \geq 0$. 

{The framework of reference \cite{dohmatob2024strong} provides a useful lens through 
which to assess the role of PPI in leveraging 
synthetic data in the context of scaling laws for AI. Their analysis shows that training on a mixture of real and 
synthetic data introduces an extra error term that scales with the distributional mismatch with real data. This term creates an 
irreducible floor in the test error that persists unless essentially all synthetic data is discarded. 
Even reweighting schemes cannot overcome this barrier, as the optimal weight 
assigned to synthetic data vanishes in the scaling-laws regime. 
PPI offers a structurally different approach: rather 
than treating synthetic and real labels as interchangeable, PPI uses a small 
real dataset to estimate and correct the systematic bias of a synthetic 
which to assess the role of PPI in leveraging 
synthetic data. The same principle underlies synthetic-powered predictive inference discussed in the next subsection.}

With $\lambda=1$, the variance of the estimator $L_{\text{PPI}}(\theta,\mathcal{D},\mathcal{D}_{\text{unl}})$ is reduced with respect to the original training loss $L_\text{train}(\theta,\mathcal{D})$ as long as the loss estimate $l=\ell(\theta,(x,\hat{y}))$ using the synthetic labels is sufficiently correlated with the true loss $\hat{l}=\ell(\theta,(x,y))$ based on the ground-truth labels   \cite{angelopoulos2023ppi,DRjordan}. Using a Gaussian approximation and denoting as $\rho$ the correlation coefficient between the variables $l$ and $\hat{l}$, this implies that the benefits of PPI grow with the mutual information $\mathrm{I}(l;\hat{l})=-\log(1-\rho^2)$.

When this mutual information decreases, by suitably setting the constant $\lambda$, one can in principle ensure that the variance of the estimate (\ref{eq:PPI}) is never larger than that of the conventional training loss based only on training data \cite{angelopoulos2024ppiplus}. In particular, setting $\lambda=0$ recovers the conventional training loss: $L_{\text{PPI}}(\theta,\mathcal{D},\mathcal{D}_{\text{unl}})=L_{\text{train}}(\theta,\mathcal{D})$. A procedure for the online optimization of the hyperparameter $\lambda$ can be found in \cite{park2025adaptivepredictionpowered} (see also \cite{einbinder2025semi}).

\subsection{Synthetic-Powered Predictive Inference}

As seen in Figure \ref{fig:3}, while PPI leverages synthetic labels to enhance parameter estimation, \textbf{synthetic-powered predictive inference (SPI)} \cite{bashari2025synthetic},  and its generalized form \textbf{general synthetic-powered predictive inference (GESPI)} \cite{bashari2025statistical},    integrate large synthetic data $\mathcal{D}_\textup{synth}$ into CP to improve the informativeness of prediction sets when the real, gold standard calibration data $\mathcal{D}_\textup{cal}$ is scarce. The synthetic data can originate from various sources, including unlabeled data with pseudo-labels, generative AI models, or auxiliary tasks.

The key theoretical guarantee of GESPI is that, under exchangeability of the calibration data, the prediction sets achieve the condition (\ref{eq:coverage}) with a slack that depends on the discrepancy between the distributions of the real and synthetic losses, as measured by total variation distance. Importantly, GESPI has an intrinsic guardrail coverage guarantee that holds true even for poor-quality synthetic data, essentially capping the slack total variation term by a user-specified threshold $\epsilon$.

Formally, let $C_{\alpha'}(x;\mathcal{D})$ denote the prediction set constructed by CP at miscoverage level $\alpha'$, based on dataset $\mathcal{D}$. GESPI constructs its prediction set as follows:
\begin{align}\label{eq:GESPI}
    &C_{\text{GESPI}}(x) = \nonumber \\  & \ \ \ C_{\alpha}(x;\mathcal{D}_\text{cal}\cup \mathcal{D}_\text{synth}) \cup C_{\alpha + \epsilon} (x;\mathcal{D}_\text{cal}) \cap  C_{\alpha}(x;\mathcal{D}_\text{cal}). 
\end{align}
Intuitively, when the synthetic and real data distributions are identical, the prediction set $C_{\alpha}(x;\mathcal{D}_\text{cal}\cup \mathcal{D}_\text{synth})$ corresponds to running CP on a larger dataset. This achieves the target miscoverage $\alpha$, while producing tighter and more stable prediction sets than standard CP.

When the synthetic data is of low quality, the GESPI prediction set (\ref{eq:GESPI}) ensures bounded miscoverage control through a two-part construction:
(\textbf{i}) it takes the union with the guardrail prediction set $C_{\alpha + \epsilon} (x;\mathcal{D}_\text{cal})$, guaranteeing that the miscoverage is upper bounded by the user-specified $\alpha+\epsilon$ level, regardless of how poor the synthetic data is; and (\textbf{ii}) it takes the intersection with standard CP set $C_{\alpha} (x;\mathcal{D}_\text{cal})$, deterministically ensuring that the prediction set is never larger than what standard CP would produce.

The benefit of GESPI lies in its improved sample efficiency: when the synthetic data $\mathcal{D}_\textup{synth}$ resemble the gold-standard real data $\mathcal{D}_\textup{cal}$, GESPI produces prediction sets with tighter coverage and size than standard CP, especially in the small-sample regime where the calibration dataset size $m$ is limited. Furthermore, regardless of the quality of the synthetic data, the size of GESPI's prediction sets remains bounded between those of standard CP and guardrail CP applied solely to $\mathcal{D}_\textup{cal}$.


\section{Conclusion and Future Directions}\label{sec:VI}

This review paper has discussed information-theoretic foundations and practical methodologies for data-efficient AI. We started by reviewing \textbf{generalized Bayesian learning} as a principled framework for modeling epistemic uncertainty in the parameter space that offers asymptotic validity properties. \textbf{Information-theoretic generalization bounds} were then reviewed as means to formalize the fundamental relationship between data availability and generalization performance, obtaining a theoretical justification for the adoption of generalized Bayesian learning. \textbf{CP} and \textbf{CRC} were presented that deliver distribution-free, finite-sample guarantees on prediction set coverage and general risk metrics, enabling reliable deployment in safety-critical applications. Finally, \textbf{PPI} and \textbf{GESPI} were shown to rigorously integrate auxiliary predictions and synthetic data with limited labeled samples, achieving more effective models and more informative prediction sets while maintaining statistical validity.

In this broad landscape, several promising research directions emerge. \textbf{Conditional coverage} remains an open challenge: while marginal coverage is well-understood, achieving valid coverage conditional on arbitrary covariate subgroups requires either strong assumptions or substantial calibration data \cite{romano2022classification}. \textbf{Epistemic uncertainty under distribution shift} needs deeper investigation, as current methods struggle to distinguish model uncertainty from environment change \cite{park2021pac}. \textbf{Scaling synthetic data methods} to settings such as reasoning in LLM \cite{jung2025prismatic} and  multimodal models, combining text, images, and sensor data, may require new principled theoretical frameworks.  Finally, thorough analyses of real-world deployments of the techniques discussed in this review for robotics, telecommunications, and healthcare would provide further evidence for the importance of uncertainty quantification and synthetic data-based AI training and calibration.

Overall, as AI systems are increasingly deployed in context-specific engineering applications, the methodologies reviewed here are envisaged to provide essential tools for responsible, data-efficient machine learning. By quantifying epistemic uncertainty and rigorously augmenting limited data, these approaches enable practitioners to navigate the fundamental tradeoff between data availability and predictive reliability.


\bibliographystyle{IEEEtran}
\bibliography{references}

\end{document}